# Metrics for Evaluating the Efficiency of Compressing Sensing Techniques

Fatima Salahdine, Elias Ghribi, Naima Kaabouch
School of Electrical Engineering and Computer Science
University of North Dakota, Grand Forks, United States
fatima.salahdine@und.edu, Elias.ghribi@und.edu, naima.kaabouch@und.edu

*Abstract*—Compressive sensing has been receiving a great deal of interest from researchers in many areas because of its ability in speeding up data acquisition. This framework allows fast signal acquisition and compression when signals are sparse in some domains. It extracts the main information from high dimensional sparse signals using only a few samples, then the sparse signals are recovered from the few measurements. There are two main points to consider when it comes to using compressive sensing. The first one is how to design the linear measurement matrix to ensure that the compressive sensing is meeting the objectives of the application. The second is how to recover the sparse signal from few measurements. Performing compressive sensing requires analyzing and investigating the efficiency of the measurement matrix and the recovery algorithm. To date, constructing explicit measurement matrices and developing efficient recovery algorithms are still open challenges in applications. Thus, this paper describes metrics to evaluate the performance of compressive sensing techniques.

## I. Introduction

Compressive sensing is a new signal acquisition and compression mechanism where the signal is captured in its compressed form [1][2]. Compressive sensing has been proposed to speed up the acquisition of high dimensional signals. It requires the signals to be sparse, which is valid in most of real-world signals. Examples of areas that apply compressive sensing include cognitive radio, medicine, magnetic resonance imaging, radar systems, and sampling theory [3-5]. In cognitive radio, compressive sensing has been applied mainly in wideband spectrum sensing to speed up the sensing and reduce the processing time. It has been also used for channel estimation in order to estimate the channel's behaviour. In medicine, compressive sensing has been considered to obtain a good image from the magnetic resonance imaging machine. It allows getting low noise images with high resolution to identify the desired pixels [4]. In radar systems, compressive sensing has been applied to reduce some targeted parameters such as: weight, data size, cost, power consumption, and complexity [5].

Compressive sensing involves three main processes, namely sparse representation, linear measurement, and recovery. The sparse representation process performs by projecting the original signal on a suitable basis where it is represented as sparse. A signal is sparse when most of its components are null. Examples of these projection basis are the fast Fourier transform, wavelet transform, and discrete cosine transform. As most of the real-world signals are sparse by nature, this process is often ignored [6]. For the rest of the paper, we assume the signals to be sparse in some domain. The linear measurements process, also called encoding, consists of multiplying the sparse signal with a measurement matrix. In this process, it is important to select a suitable matrix that best represents the sparse signal allowing efficient recovery of the original signal. In order to ensure the convenience and the relevance of the measurement matrix and thereby guarantee a successful recovery, two requirements need to be met, namely satisfying the restricted isometry property (RIP) and the coherence by the measurement matrix [7]. The recovery process, also called decoding, consists of recovering the original sparse signal from few measurements at the receiver.

The efficiency evaluation of compressive sensing in meeting the objectives depends on evaluating both the measurement matrix and the recovery algorithm. A number of evaluation metrics have been used over the literature. For instance, in [3], the authors used processing time as metric to evaluate the speed of the measurement matrix for real time spectrum sensing. In [8], the authors used the failure rate as an evaluation metric to evaluate how accurate is a recovery algorithm in reconstructing high dimensional signals. In [9], the authors compared the efficiency of their recovery algorithm using the mean square error. In [10], the authors compared the performance of the sampling matrices using metrics, namely recovery error, processing time, recovery time, covariance, and phase transition diagram. In [11], the authors evaluated the compressive sensing technique based on the recovery success rate, reconstruction error, recovery time, compression ratio, and processing time. In [12], the authors investigated the sparsity of the Bayesian compressive sensing based on the error sparsity and signal sparsity.

Moreover, the authors in [13] used several metrics to evaluate the efficiency of the compressive sensing based Circulant matrix and Bayesian recovery in cognitive radio networks. These metrics are mean square error, recovery error, sampling time, recovery time, processing time, and correlation coefficients. In [14], the authors investigated the accuracy of the normalized iterative hard thresholding recovery algorithm using processing time, recovery time, recovery error, required number of measurements, and recovery success rate. In [15], the authors compared the results of the compressive sensing based Toeplitz matrix and Bayesian recovery with those of the existing works using: sparsity, sampling time, recovery time, processing time, recovery error, mean square error, signal to noise ratio, and required number of measurements. In [16], other metrics were used to evaluate the Toeplitz sampling matrix, including the measurements' number, number of coefficients of the partial Toeplitz matrix, complexity, storage cost, and probability of returning the true solution. In [17], the authors compared one-bit compressive sensing to multi-bit compressive sensing using hamming distance, recovery signal to noise ratio, complexity, number of required measurements, recovery error, and processing time. Some of these metrics have similar similarities.

In this paper, we focus on practical methodologies to achieve a good evaluation of compressive sensing processes in any application. Thus, we represent an in depth survey on different evaluation methods and metrics that can be used by researchers to investigate the efficiency of their compressive sensing techniques. The remaining of this paper is organized as follows: Section II reviews the compressive sensing theory. Section III describes practical evaluation methodologies of the encoding and decoding processes and discusses the most relevant evaluation metrics used by several applications. In Section IV, challenges and future research works are discussed. Finally, a conclusion is given at the end.

## II. COMPRESSIVE SENSING THEORY

Compressive sensing requires that the original signals are sparse in some domains. A signal, $x$, is sparse when most of its values are null. The measurement matrix process consists of multiplying the sparse signal, x, of $N$ coefficients with a measurement matrix, $A$, of $M$x$N$ elements where $M<<N$. This multiplication extracts the main information from the original sparse signal and removes the rest. The output signal, $y$, is the measurements signal of $M$ coefficients. The mathematical model of the compressive sensing is expressed as

$$y = Ax \quad (1)$$

where $x$ denotes the sparse signal, $A$ denotes the measurement matrix, and $y$ denotes the signal measurements [18]. In the presence of the noise, equation (1) can be rewritten as

$$y = Ax + n \quad (2)$$

where $n$ represents the noise [19]. Fig. 1 illustrates how the signal acquisition is performed by taking only a few measurements of the original signal at the receiver.

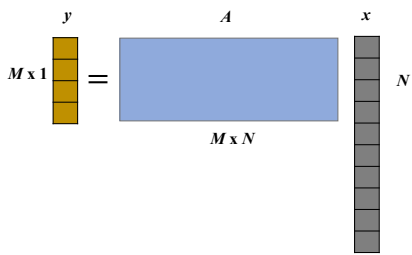

Fig. 1. General model of compressive sensing.

Measurement matrix is used to simultaneously sample and compress the original sparse signal below the Nyquist rate. It allows reducing the number of captured samples while retaining the important information from the original signal. Examples of the measurement matrices are: random matrix, Gaussian matrix, Bernoulli matrix, Deterministic matrix, Circulant matrices, and Toeplitz matrices [6][20].

At the receiver, the original signal can be recovered from the signal measurements by solving the underdetermined system of linear equation given as

$$\underset{x}{\text{minimize}} \; \|x\|_0 \; \text{subject to} \; y = Ax \quad (3)$$

where $\|.\|_0$ is the $l_0$-norm, $x$ is the sparse signal, $A$ is the measurement matrix, and $y$ is the signal measurements. As solving this system is challenging, its convex relaxation is considered by replacing the $l_0$-norm with $l_1$-norm [21]. The recovery problem can be then re-written as

$$\underset{x}{\text{minimize}} \; \|x\|_1 \; \text{subject to} \; y = Ax \quad (4)$$

where $\|.\|_1$ denotes the $l_1$-norm. Recovery algorithms, also called reconstruction algorithms, are used to recover the original sparse signal from few measurements. They perform by solving the underdetermined system presented in equation (3) as an optimization problem. They are classified into three main categories: convex relaxation, Greedy, and Bayesian based recovery [22]. Convex relaxation algorithms perform by solving the equation (3) using linear programming. Basis pursuit and iterative thresholding are examples of algorithms classified under the convex relaxation category [23]. Greedy algorithms perform by selecting iteratively a local optimal corresponding to the position of a non-zero coefficient of the signal until estimating the whole signal. Matching pursuit and orthogonal matching pursuit are examples of algorithms classified under the Greedy category [24]. Bayesian based recovery algorithms perform by estimating the unknown coefficients through probabilities and the known evidences. They consider estimating prior probability distribution in order to find the full posterior probability distribution of the unknown signal's coefficients. Bayesian based belief propagation and Laplace priors are examples of algorithms classified under the Bayesian based recovery category [11].

## III. EVALUATION METRICS

Applying compressive sensing in solving any problem involves investigating the efficiency and the relevance of the measurement matrices as well as the recovery algorithms. A number of evaluation metrics have been proposed and used in different applications [25]. These metrics include coherence, sparsity, recovery error, correlation, recovery time, processing time, compression ratio, and phase transition diagram. They cover the most important aspects of the three compressive sensing processes in terms of time, error, rate, and cost.

### 1. Coherence

Coherence metric evaluates the quality of the measurement matrix and guarantees the success of the recovery process. It refers to the maximum value of correlation between two normalized columns of the sampling matrix [26]. Let $A$ denotes a matrix with $l_2$-normalized columns, i.e., $\|a_i\|_2 = 1$, $i = 1, ..., N$. The coherence, $\mu$, of the matrix $A$ is defined as

$$\mu := \underset{1 < i \neq j < N}{\max} | < a_i, a_j > | \quad (5)$$

where $a_i$ and $a_j$ denote two columns of the matrix $A$ and $N$ denotes the number of samples. Low coherence implies few measurements are required to reconstruct the original signal. The smaller the coherence, the better the reconstruction algorithm performs. However, the lower bound on the coherence limits the recovery algorithms performance to rather small sparsity levels. Moreover, the coherence, $\mu$, of a matrix, $A$, of $M$x$N$ satisfies the following condition

$$\mu \geq \sqrt{\frac{N-M}{M(N-1)}} \quad (6)$$

where $N$ and $M$ are the number of samples and the number of measurements respectively. This equality holds true if and only if the columns $a_1, \ldots, a_N$ of the matrix, $A$, form an equiangular tight frame [6].

2. *Null space property*

The null space of the matrix, $A$, is denoted as

$$\aleph(A) = \{x : Ax = 0\} \quad (7)$$

In order to recover the original sparse signal from the signal measurements, $y = Ax$, any pair of distinct vectors $x$, $x' \in \sum_{2k}$, are required to satisfy the following condition

$$Ax \neq Ax' \quad (8)$$

This condition is called null space property (NSP) [25]. In case this condition is not satisfied, it would be impossible to recover the signal, $x$, and distinguish it from $x'$ based only on the signal measurements, $y$. This condition implies that if $Ax = Ax'$, then $A(x - x') = 0$ with $x - x' \in \sum_{2k}$, which gives that $A$ represents all $x \in \sum_k$ if and only if the null space of the matrix, $\aleph(A)$, do not contain any vector of $\sum_{2k}$. In order to clarify this condition, matrix spark is considered. The matrix spark of a matrix, $A$, is defined as the smallest number of linearly independent columns. It is computed by checking dependence of columns combination. The null space condition can be formulated using the spark. For any signal measurements, $y$, there is at most one signal, $x$, that satisfies the following condition

$$x \in \sum_k, y = Ax \text{ if and only if } \operatorname{spark}(A) > 2k \quad (9)$$

where $k$ denotes the order. For exact sparse signals, matrix spark condition implies that recovering the sparse signals is possible. For signals approximately sparse, more conditions should be considered. Null space condition metric provides accurate performance evaluation of the sampling matrix. However, it is highly expensive and not practical [25].

3. *Restricted Isometry Property*

Restricted isometry property characterizes the orthonormal matrices, which are bounded with a constant called restricted isometry constant [27]. A sampling matrix, $A$, is satisfying the restricted isometry property of order $k$ when the following condition is true

$$\exists \delta_k \in (0,1) / (1 - \delta_k) \|x\|_2^2 \leq \|Ax\|_2^2 \leq (1 + \delta_k) \|x\|_2^2 \quad (10)$$

where $\|.\|_2$ is the $l_2$-norm, $\varepsilon_k \in [0,1]$ is the restricted isometry constant of the matrix $A$, and $x$ is the $k$-sparse signal [6]. Restricted isometry property must be considered when designing the sampling matrix. It is required by many recovery algorithms as it guarantees the uniqueness of the solution of the underdetermined system, which represents the reconstructed signal. Examples of matrices satisfying this condition are random and Gaussian matrices [27].

4. *Relationship between NSP and RIP*

When a matrix satisfies the restricted isometry property, it implies that the matrix also satisfies the null space property [26]. Let consider a sampling matrix, $A$, satisfying the RIP of order $2k$ where $\delta_{2k} < \sqrt{2} - 1$, thus, $A$ satisfies the NSP of order $2k$ with constant equal to

$$C = \frac{2}{1 - (1 + \sqrt{2})k} \quad (11)$$

where $C$ denotes the NSP constant. One can conclude that the RIP is strictly stronger than the NSP.

5. *Sparsity level*

The sparsity of the signal is required in order to perform compressive sensing. Sparsity feature characterizes most of real world signals [21]. It consists of projecting a signal on a sparse basis where most of its values are zeros. A signal, $x$, of $N$ samples is sparse of order k when it has only $k$ non-zero coefficients and $k \ll (N-k)$. Sparsity feature is also referred to the universality, which implies that the original signal can be represented as sparse in any basis or domain [28]. In addition, non-sparse signals can be also projected on specific basis by using some sparse representation [6].

6. *Error sparsity*

Error sparsity metric investigates the sparsity level of the recovered signal after the recovery algorithm [2]. It performs by comparing the sparsity of the original signal with the sparsity of the estimated signal to obtain the error sparsity.

7. *Measurement bounds*

The measurements bounds metric refers to how many measurements are required to satisfy the restricted isometry property while ignoring the impact of its constant $\delta_{2k}$ [20]. Let consider an $s$-sparse signal of $N$ samples and an $M$x$N$ measurement matrix, $A$, that satisfies the RIP of order $2k$ with a constant $\delta_{2s} \in [0, 0.5]$. The number of measurements, $M$, required to satisfy the RIP of order $2k$ is given as

$$M \geq C \, s \, \log\left(\frac{N}{k}\right) \quad (12)$$

where $C$ is a positive constant approximately, $N$ is the number of samples of the sparse signal, and $s$ denotes the sparsity level.

The required number of measurements refers to the minimum number of measurements that can be extracted from the original signal while compression [6]. It must be sufficient to recover all the coefficients of the original signal from it. It is similar to the measurements bounds as each sampling matrix requires a specific measurements to perform well. For instance, for a random sampling matrix, the number of measurements, $M$, required is given by

$$M = O(\log N / s) \quad (13)$$

where $N$ is the number of samples and s is the sparsity level.

8. *Recovery error and Mean square error*

Recovery error, also called reconstruction error, is the norm of the difference between the original signal and the recovered signal divided by the original signal's norm [13][15]. it can be expressed as

$$R = \frac{\|x - \hat{x}\|}{\|x\|} \quad (14)$$

where $R$ is the recovery error, $x$ is the original signal, and $\hat{x}$ is the recovered signal. It represents the error level of a recovery algorithm. Another metric that gives similar results is Mean square error, *MSE*. This metric is used to evaluate the variation of the recovery error of a recovery algorithm over time. It is mainly used for predictive modeling about a recovery algorithm for error estimation [15]. It measures the average magnitude of the squared difference between the original signal and the recovered signal. It can be expressed as

$$MSE = \frac{\sum_n [x(N) - \hat{x}(N)]^2}{N} \quad (15)$$

where *MSE* denotes the mean square error, $x$ denotes the original signal, $\hat{x}$ denotes the recovered signal, and $N$ denotes the number of samples. Mean square error refers to the amount of error by which the recovered signal and the original signal are different. Both recovery error and mean square error are almost similar.

## 9. Correlation and covariance

Correlation, *C*, measures the similarities between the recovered signal and the original signal. It measures how similar the two signals are and it is between -1 and 1 [13]. It is expressed as

$$C = \frac{N \sum(x\tilde{x}) - (\sum x)(\sum \widetilde{x})}{\sqrt{N(\sum x^2) - (\sum x)^2}\sqrt{N(\sum \tilde{x}^2) - (\sum \tilde{x})^2}} \quad (16)$$

Positive correlation implies that the original signal and the recovered signal are positively correlated. Negative correlation implies that the original signal and the recovered signal are negatively correlated. Null correlation implies no relationship between the two signals.

Covariance metric is a statistical measure corresponding to the correlation between the recovered signal and the original signal. It can be defined as

$$cov = E([x - E(x)][\hat{x} - E(\hat{x})]) \quad (17)$$

where $E(.)$ denotes expectation, $x$ denotes the original signal, and $\hat{x}$ denotes the reconstructed signal. Both correlation and variance measure the dependency and the relationship between the original signal and the recovered. Despite they have some similarities and they can be considered as one metric, they are different.

## 10. Sampling time

Sampling time metric measures the amount of time required by the sampling matrix process to acquire and compress the original signal using a sampling matrix [13-15]. It evaluates the sampling speed of a measurement matrix in order to define the fastest sampling technique.

## 11. Recovery time

Recovery time metric measures the time required by a recovery algorithm to solve the sparse recovery problem. It examines how fast is a recovery algorithm [13-16]. Moreover, we refer to the processing time when considering all the compressive sensing processes to evaluates how fast is a compressive sensing technique.

## 12. Compression Ratio

Compression ratio, *CR*, is the ratio between the number of measurements and the number of samples in the original signal [6]. It can be computed as

$$CR = \frac{M}{N} \quad (18)$$

where *M* is the number of measurements and *N* is the number of samples. This metric verifies that high dimensional signals can be recovered using few measurements.

## 13. Signal to error ratio

Signal to error ratio, also called signal to noise ratio, measures the strength of a signal over the noise [15-17]. In compressive sensing, this metric measures the strength of the original signal over the recovered signal. It can be defined as

$$SNR = 10 \log_{10} \frac{\sum_N [x(N)]^2}{\sum_N [x(N) - \hat{x}(N)]^2} \quad (19)$$

where SNR is the signal to noise ratio. Indeed, signal to error ratio can also refer to the recovery SNR, which represents the SNR of the recovered signal. Recovery SNR aims to verify the noise level in the recovered signal as a recovery algorithm can recover only noise.

## 14. Recovery signal to noise ratio

Recovery signal to noise ratio metric measures the *SNR* level at the receiver by considering the original signal the input and the recovered signal the output [12-16]. It is expressed as

$$RSNR = \frac{\|x\|_2^2}{E(\|x - \hat{x}\|_2^2)} \quad (20)$$

where *RSNR* denotes the recovery signal to noise ratio.

## 15. Recovery success rate and failure rate

Recovery success rate is the success rate of a recovery algorithm indicating how successful an algorithm is. It counts the number of times the original signal and the recovered signals are almost similar (90%) for different values of sparsity level, number of samples, and number of measurements [8][11].

Failure rate is a different metric; but gives the same information. Failure rate is the success rate reverse and it is computed over a number of experiments to count the number of times the recovery algorithm fail to recover the original signal [11].

## 16. Phase Transition diagram

Phase transition diagram determines the recovery success of a recovery algorithm by representing the probability of recovery success against the probability of recovery failure [20]. This metric evaluates the performance of both measurement matrix and recovery processes. A representation of success area and failure area can be considered in a phase space of the pair $(\rho, \delta)$, where $\delta=M/N$ denotes the compression ratio and $\rho=K/M$ denotes the ratio of the signal sparsity and the number of measurements.

Phase transition diagram provides the relationship between the most important parameters of a compressive sensing technique, namely sparsity level, number of samples, and number of measurements. It plots these parameters to separate success phase from failure phase of a given algorithm. It helps determining which values to select in order to occur success phase by the measurement matrix design and the recovery.

## 17. Hamming distance

Hamming distance determines the number of times the original signal and the recovered signal are different [17]. It is the number of non-zero coefficients of *H* where $H = y - \hat{y}$, y is the noisy measurements, and $\hat{y}$ is the noisy recovered signal. Its minimum corresponds to accurate signal recovery.

## 18. Complexity

Complexity reflects the algorithm's efficiency to perform with large amount of data. It can be computational complexity, time complexity, or hardware complexity. In compressive sensing, the complexity of designing a sampling matrix, acquiring a high dimensional signal, or performing a recovery process must be considered [16]. Complexity depends on sparsity, number of samples, and number of measurements.

IV. DISCUSSION

The evaluation metrics allow investigating the efficiency of each compressive sensing process. Each metric corresponds to one or more processes. Table I classifies these metrics according to which process is involved.

TABLE I. EVALUATION METRICS

| Metrics | Sparse Representation | Sampling matrix | Recovery |
|---|---|---|---|
| Coherence | | x | x |
| RIP | | x | |
| NSP | | | x |
| Sparsity | x | | |
| Error sparsity | x | | |
| Measurements bounds | | x | x |
| Recovery error, MSE | | | x |
| Correlation/covariance | | | x |
| Recovery time | | | x |
| Sampling time | | x | |
| Compression ratio | | x | x |
| Signal to error ratio | | x | |
| Recovery SNR | | x | |
| Recovery success rate/ Failure rate | | x | |
| Phase transmission diagram | | x | |
| Recovered SNR | | x | |
| Hamming distance | | x | |
| Complexity | x | x | x |

The evaluation of the performance of each compressive sensing process can be performed using one or several metrics shown in Table I. Computing these metrics requires performing many experiments in order to verify the accuracy of the compressive techniques. It is actually a challenge to select which best performance metric can evaluate a compressive sensing approach for a given application. The choice of which metric to adopt depends mainly on the application objectives. When looking for a fast acquisition technique, sampling time can be considered, but recovery accuracy may not be achieved [6]. For instance, random matrices are slow but easy to implement. Toeplitz and Circulant matrices are able to reduce the randomness compared to others. Gaussian and Bernoulli matrices are also simple to implement, but their hardware implementation is expensive. Deterministic matrices require the number of measurements to be more than the expected threshold, but random convolution matrices require less measurements. Thus, sampling time metric is not sufficient to evaluate a matrix performance, other metrics must be considered simultaneously, including complexity, randomness level, and measurements number.

Fast compressive sensing techniques are not always efficient in terms of recovery error, and low error and accurate techniques are not always fast [29]. Thus, there is a trade-off between the different performance metrics. For instance, null space property guarantees high recovery rate, but it does not take into consideration the noise. Coherence limits the performance evaluation of the recovery algorithms to rather small sparsity levels [30]. Complexity of an algorithm limits its performance in terms of processing time and cost. Costly algorithms may not be considered even if they provide accurate results [31].

EXAMPLES OF APPLICATIONS

In order to evaluate the efficiency of the compressive sensing techniques, adequate and specific evaluation metrics should be used and quantified. In general, signal sampling requires robust and structured matrices, which can be evaluated by using sampling time and complexity. Signal recovery requires accurate and fast reconstruction algorithms, which can be evaluated using the recovery error, recovery time, recovery complexity, and hardware cost. Which evaluation metrics should be used depends mainly on the application and its objectives. Below are a few examples of areas applying compressing sensing along with the appropriate evaluation metrics.

*1. Wideband Spectrum Sensing*

In cognitive radio, secondary users (SUs) can sense the radio spectrum to identify free licensed channels and use them under the condition to not create any harmful interference to primary users' signals. Compressive sensing can be used to speed up the spectrum sensing process which allows SUs to access the free channels before they get occupied by their owners. In this case, sensing time or processing time is one of the best metrics to be used to evaluate the speed of the spectrum scanning process. Other important metrics to be used in this application include the probability of detection, the probability of false alarm, and the probability of miss-detection should be used [32][33].

*2. Cooperative sensing*

Compressive sensing has also been proposed to speed up the process in cooperative sensing. In this process, several SUs sense the radio spectrum and send their sensing reports to a fusion center. This center compiles these reports and take the final sensing decision on each frequency channel and then send the decisions to all SUs [34]. As the number of the cooperating SUs increases, the time needed by the fusion center to process the reports, take the sensing decision, and send the final sending decision to SUs increases. However, the accuracy of the final sensing decision increases with the increase of the cooperating SUs. Thus, appropriate metrics to evaluate compressive sensing techniques for this application include the processing time taken by the SUs and the fusion center, the probability of detection, the probability of false alarm, the probability of miss-detection, recovery rate, and complexity.

*3. Sensors networks*

Wireless sensors network is another area for which compressive sensing has been proposed. The efficiency evaluation of compressive sensing techniques in this area should include several metrics, including transmission power, bandwidth, sparsity, and energy consumption [35]. The energy consumption is considered another important metric to use when applying compressive sensing techniques in wireless sensor networks as these techniques can minimize the energy while achieving high performance.

*4. Biomedical Signals/Images*

Compressive sensing has also been proposed for biomedical signaling/imaging devices in order to reduce the sampling load required by the biomedical sensors [36]. Compressive sensing allows speeding up the sparse biomedical signals/images scanning and processing via low power medical sensors. As biomedical signals/images have to be accurate for detecting abnormalities, metrics such as recovery error, recovery success, recovery SNR, correlation should be used to compare the efficiency of techniques with and without compressive sensing. In addition, as the lifetime of a biomedical sensor depends mainly on the power consumption, this metric can be used for the performance evaluation as well. Other metrics that can be considered include sampling load, cost, transmission capacity, storage usage, and transmission time.

## Conclusion

Compressive sensing allows acquiring the main information from high dimensional sparse signals by using a sampling matrix and then a recovery algorithm to reconstruct the original signal. Designing accurate and fast sampling matrices is important in terms of guaranteeing the signal's recovery at the receiver. Developing recovery algorithm is also important to obtain recovered signal with high success rate and less recovery time. In this paper, we reviewed the different performance metrics used to evaluate the efficiency of the compressive sensing processes. We compared these metrics in terms of which process is involved. Each process can be evaluated using a number of metrics in order to quantify its accuracy. Deciding which metric to adopt among several is still challenging the compressive sensing in many applications. As this choice depends on the application's interest: fast or accurate results.